\documentclass[preprint, aps, preprintnumbers, superscriptaddress, nofootinbib, floatfix]{revtex4-1}

\usepackage{amsmath}
\usepackage{bm}
\usepackage{amsfonts}
\usepackage{graphicx}
\usepackage{hyperref}
\usepackage{color}
\usepackage{ulem}

\hypersetup{
colorlinks=true,
linkcolor=blue,
linktoc=page,
citecolor=blue,
urlcolor=blue}

\begin{document}

\title{\bf The Borde-Guth-Vilenkin Theorem in extended de Sitter spaces}

\newcommand{\FIRSTAFF}{\affiliation{Department of Physics, University at Buffalo, Buffalo, NY 14260, USA}}
\newcommand{\SECONDAFF}{\affiliation{Centre for Strings, Gravitation and Cosmology, Department of Physics, Indian Institute of Technology Madras, Chennai 600 036, India}}

\author{William H. Kinney}
\email[Electronic address: ]{whkinney@buffalo.edu}
\FIRSTAFF
\author{Suvashis Maity}
\email[Electronic address: ]{suvashis@physics.iitm.ac.in}
\SECONDAFF
\author{L. Sriramkumar}
\email[Electronic address: ]{sriram@physics.iitm.ac.in}
\SECONDAFF
\date{\today}
\begin{abstract}
The Borde-Guth-Vilenkin (BGV) theorem states that any spacetime with net positive expansion must be geodesically incomplete. 
We derive a new version of the theorem using the fluid flow formalism of General Relativity. The theorem is purely kinematic, depending on the local expansion properties of geodesics, and makes no assumptions about energy conditions. We discuss the physical interpretation of this result in terms of coordinate patches on de Sitter space, and apply the theorem to Penrose's model of Conformal Cyclic Cosmology. We argue that the Conformal Cyclic extension of an asymptotically de Sitter universe is geodesically incomplete. 
\end{abstract}

\maketitle

\section{Introduction}

The geodesic incompleteness theorem of Borde, Guth, and Vilenkin \cite{Borde:2001nh} is a general and powerful constraint on the structure of cosmological spacetimes. A \textit{geodesically complete} spacetime is one for which all possible inertial paths in the spacetime extend infinitely in proper time toward both the past and the future. By contrast, a geodesically \textit{past-incomplete} spacetime contains geodesics which encounter a boundary at finite past proper time, and a geodesically \textit{future-incomplete} spacetime contains geodesics which encounter a boundary at finite future proper time.  The Borde-Guth-Vilenkin (BGV) Theorem states that any spacetime for which the net expansion is positive,
\begin{equation}
    \int{H dt} > 0,
\end{equation}
is necessarily geodesically past-incomplete. This has profound implications for cosmological model building, since it applies not only to trivially past-incomplete spacetimes such as matter- or radiation-dominated Friedmann-Robertson-Walker (FRW) cosmologies, but to inflationary models as well. Furthermore, the proof is purely kinematic, and does not depend any assumption about energy conditions. Recently proposed cyclic cosmologies, which attempt to remove the need for an initial cosmological singularity using alternating periods of expansion and contraction, have also been shown to be past-incomplete \cite{Kinney:2021imp}. 

In this paper, we derive a version of the BGV Theorem using the fully covariant, coordinate-independent fluid flow formalism for General Relativity, based on the foliation of spacetime into hypersurfaces orthogonal to a timelike congruence of geodesic four-velocity vectors $\left\lbrace u^\mu\right\rbrace$. In this formalism, the Hubble constant $H$ is generalized to the fluid four-divergence, $\Theta \equiv u^\mu{}_{;\mu}$, and the scale factor is generalized to the components of the orthogonal projection tensor $\lambda_{\mu \nu}:\ \lambda_{\mu \nu} u^{\nu} = 0$. We show that any local spacetime interval for which the net expansion integrated over the proper time $ds$ along a geodesic is positive,
\begin{equation}
    \int{\Theta ds} > 0,
\end{equation}
is geodesically past-incomplete, and any region for which the net expansion is negative, 
\begin{equation}
    \int{\Theta ds} < 0,
\end{equation}
is geodesically future-incomplete. The proof is independent of choice of metric and energy conditions. Our proof differs from the proof in Ref. \cite{Borde:2001nh} in that it uses locally defined geometric variables. We emphasize that this is no \textit{more} general than the original proof, but formulates the theorem in a new way.

The paper is organized as follows: In Sec. \ref{sec:BasicBGVTheorem}, we review the BGV Theorem in the simple case of FRW cosmology. In Sec. \ref{sec:BGVGeneral} we present the geometric proof of the BGV Theorem using the fluid flow formalism. In Sec. \ref{sec:Raychaudhuri}, we include spacetime dynamics and discuss the geodesic structure of de Sitter spacetimes, identifying the incomplete geodesics in the BGV construction with rest-frame geodesics in the open, $k = -1$ patch of the de Sitter manifold. Finally, we apply the results to the Conformal Cyclic extension of an asytmpotically de Sitter spacetime, and argue that such extended spacetimes are geodesically incomplete, independent of the details of the conformal matching at the boundary between cosmological ``aeons''. Section \ref{sec:Conclusions} contains discussion and conclusions. 

\section{The BGV Theorem in Friedmann-Robertson-Walker space}
\label{sec:BasicBGVTheorem}

Before constructing a more general proof, we begin by reviewing the BGV theorem \cite{Borde:2001nh} for geodesic incompleteness in an FRW spacetime. We take a metric of the form
\begin{equation}
    ds^2 = - dt^2 + a^2\left(t\right) d \mathbf{x}^2,
\end{equation}
where we assume a flat spatial geometry, $d \mathbf{x}^2 = \delta_{i j} dx^i dx^j$, for convenience. We wish to address the question of whether such spacetimes are \textit{geodesically complete}, which is the statement that all inertial (or \textit{geodesic}) observers see the spacetime as both past- and future-infinite, as measured by the proper time along a geodesic world line in the spacetime. 
\begin{equation}
    \int_{-\infty}^t{d s} \rightarrow \infty, \qquad \int_t^{+\infty}{ds} \rightarrow \infty. 
\end{equation}
The metric is singular for $a\left(t\right) = 0$, so that standard matter- or radiation-dominated cosmologies, with $a\left(t\right) \propto t^\alpha$, are trivially geodesically incomplete, since $a\left(t\right) \rightarrow 0$ at finite past time. The case of de Sitter space, 
\begin{equation}
    a\left(t\right) \propto e^{H t},\qquad H = \mathrm{const.},
\end{equation}
is more ambiguous, since the initial singularity is reached only at $t \rightarrow -\infty$ for a comoving observer. It is, however, straightforward to show that non-comoving geodesics are past-finite. For a timelike geodesic, normalization $u^\mu u_\mu = -1$ gives
\begin{equation}
    u^\mu u_\mu = g_{\mu \nu} \frac{d x^\mu}{ds} \frac{d x^\nu}{ds} = - \left(\frac{d t}{ds}\right)^2 + a^{2}\left(t\right) \left\vert \frac{d \mathbf{x}}{d s}\right\vert^2 = -1.
\end{equation}
The geodesic equation for the motion free falling observer is
\begin{equation}
    \frac{d}{d s}\left[a^2\left(t\right) \frac{d \mathbf{x}}{d s}\right] = 0,
\end{equation}
so that we can define an integration constant $v_0$ such that \cite{Aguirre:2001ks}
\begin{equation}
a^2\left(t\right) \left\vert \frac{d \mathbf{x}}{d s} \right\vert \equiv v_0 = \mathrm{const.}
\end{equation}
We then have
\begin{equation}
    \left(\frac{d t}{ds}\right)^2 = \gamma^2 = 1 + v_0^2 a^{-2}\left(t\right),
    \label{eq:dtdstimelike}
\end{equation}
where
\begin{equation}
    \gamma^2 = \frac{1}{1 - \mathbf{v}^2}
\end{equation}
is the Lorentz boost as a function of the three-velocity $\mathbf{v}$ of the observer, so that the integration constant $v_0$ is
\begin{equation}
    v_0^2 = \frac{a^2 \mathbf{v}^2}{1 - \mathbf{v}^2} = \mathrm{const.}
\end{equation}
The differential proper time $ds$ along the geodesic of the freely falling observer can then be written in terms of the coordinate time $dt$ as 
\begin{equation}
    ds = \frac{dt}{\sqrt{1 + v_0^2 a^{-2}\left(t\right)}}. 
\end{equation}
For $a\left(t\right) = e^{H t}$, the integral
\begin{align}
    \Delta s &= \int_{-\infty}^{0}{\frac{dt}{\sqrt{1 + v_0^2 e^{-2 H t}}}}\\
    &= \frac{1}{2 H} \ln \left(\frac{\sqrt{1 + v_0^2} + 1}{\sqrt{1 + v_0^2} - 1}\right)\\
    &= \frac{1}{2 H} \ln \left(\frac{\gamma_0 + 1}{\gamma_0 - 1}\right),
    \label{eq:DeltasdeSitter}
\end{align}
where the Lorentz factor $\gamma_0 = \sqrt{1 + v_0^2}$ is evaluated at $t = 0$. This integral is finite for any $v_0 \neq 0$, so the observer on a non-comoving world line sees the space as past-finite, and the space is geodesically past-incomplete. 

The BGV theorem states that any spacetime for which the average expansion rate $\mathcal{H}_{\mathrm av}$ is positive is geodesically incomplete, where
\begin{equation}
    \mathcal{H}_{\mathrm av} \equiv \frac{1}{\Delta s} \int{H ds},
    \label{eq:Hav}
\end{equation}
and $ds$ is the proper time along a geodesic world line. First, let us consider the integral of the expansion rate $H$ along a comoving world line, for which the proper time $ds = dt$,
\begin{align}
    \int_{t_i}^{t_f}{H dt} &= \int_{t_i}^{t_f}{\frac{1}{a} \frac{d a}{d t} dt}\\
    &= \int_{a_i}^{a_f}{d \ln{a}} = \ln{\left(a_f\right)} - \ln{\left(a_i\right)}. 
    \label{eq:Hdt}
\end{align}
Since the integrand is a total derivative, the integral is determined by the the values of the scale factor on the boundary, and measures the total expansion from $t_i$ to $t_f$. Note in particular that this integral is logarithmically divergent as $a_i \rightarrow 0$, consistent with the fact that comoving world lines are past-infinite in proper time. (We will generalize this expression to non-FRW metrics in Sec. \ref{sec:BGVGeneral}.) We can integrate the expansion rate $H$ along a non-comoving geodesic with proper time $d s$ as follows:
\begin{align}
    \int_{s_i}^{s_f} H ds &= \int_{t_i}^{t_f} H \frac{d s}{d t} dt \\
    &= \int_{t_i}^{t_f}{\frac{1}{a}\left(\frac{d a}{d t}\right) \frac{1}{\sqrt{1 + v_0^2 / a^2}} dt} \\
    &= \int_{a_i}^{a_f}{\frac{d a}{\sqrt{a^2 + v_0^2}}} \\
    &= \frac{1}{2} \left.\ln{\left[\frac{\sqrt{v_0^2 + a^2} + a}{\sqrt{v_0^2 + a^2} - a}\right]} \right\vert_{a_i}^{a_f}. 
\end{align}
Integrating from $a_i = 0$ to $a_f = 1$, this integral is finite:
\begin{align}
    \int_{s_i}^{s_f} H ds &= \frac{1}{2} \ln \left(\frac{\sqrt{1 + v_0^2} + 1}{\sqrt{1 + v_0^2} - 1}\right)\\
    &= \frac{1}{2} \ln \left(\frac{\gamma_0 + 1}{\gamma_0 - 1}\right).
    \label{eq:BGVRelation}
\end{align}
Note that unlike the case of Eq. (\ref{eq:DeltasdeSitter}), we have not taken $H = \mathrm{const.}$ The only assumption is that the overall change in scale factor is positive, $a_f > a_i$. We can then define an average expansion rate
\begin{equation}
    \mathcal{H}_\mathrm{av} \equiv \frac{1}{\Delta s} \int{H ds}. 
\end{equation}
The BGV Theorem is then the statement that as long as $\mathcal{H}_\mathrm{av}$ is finite and positive, the proper time $\Delta s$ along a timelike worldine is finite
\begin{equation}
    \Delta s = \int{d s} = \frac{1}{\mathcal{H}_\mathrm{av}} \int{H ds}. 
    \label{eq:BGV}
\end{equation}

The BGV Theorem as stated in Eq. (\ref{eq:BGV}) is, however, something of a circular argument, since $\mathcal{H}_\mathrm{av}$ is only well-defined if $\Delta s$ is finite in the first place. A better way to state it is that $\mathcal{H}_\mathrm{av}$ depends only on the \textit{amount} of expansion, not the specific expansion history, as can be seen from Eq. (\ref{eq:Hdt}). For a given expansion history $H\left(s\right)$, we can define a bounding de Sitter space with $H  = \mathcal{H}_\mathrm{av} = \mathrm{const.}$ such that 
\begin{equation}
    \mathcal{H}_\mathrm{av} \int_{s_i}^{s_f} ds = \int_{s_i}^{s_f} H ds, 
\end{equation}
and the ratio $a_f / a_i$ on the interval $[s_i, s_f]$ is the same for both spaces. We can then choose a scale factor $a_0$ in the bounding de Sitter space such that
\begin{equation}
    a_0 e^{\mathcal{H}_{\mathrm{av}} t} \geq a(t),\quad \forall t \in [t_i,t_f].
\end{equation}
The past proper time along a geodesic in the target space is then bounded from above by the past proper time in the bounding de Sitter space \cite{Kinney:2021imp},
\begin{equation}
    \Delta s = \int_{t_i}^{t_f}{\frac{d t}{\sqrt{1 + v_0^2 a^{-2}\left(t\right)}}} \leq \int_{t_i}^{t_f}{\frac{d t}{\sqrt{1 + v_0^2 a_0^{-2} e^{-2 \mathcal{H}_{\mathrm{av}} t}}}}.
\end{equation}
Then geodesic incompleteness of the bounding de Sitter space implies geodesic incompleteness of the space with expansion history $H\left(s\right)$, which we will refer to in what follows as the target space. This applies not only to spaces where the scale factor $a$ grows monotonically, but also to cyclic universes for which the scale factor $a\left(t\right)$ is oscillatory, but grows from one cycle to the next \cite{Kinney:2021imp}. We have so far only demonstrated this for an FRW metric; in the next section we generalize this to non-FRW spacetime. 

\section{A geometric formulation of the BGV Theorem}
\label{sec:BGVGeneral}

\begin{figure*}
\centerline{\includegraphics[width=4.5in]{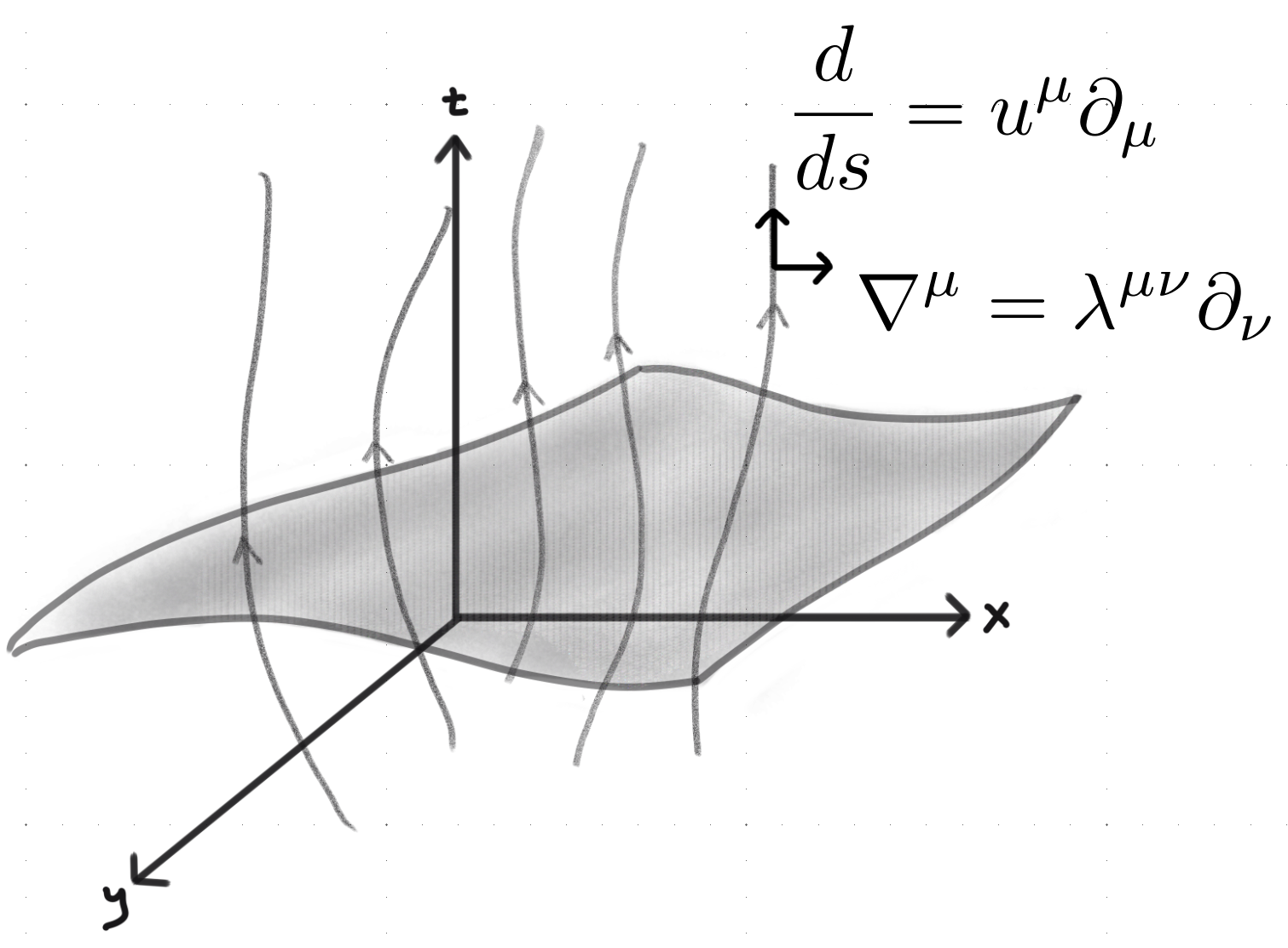}}  
\caption{A foliation of spacetime into spacelike hypersurfaces orthogonal to a timelike congruence $\left\lbrace u^\mu\right\rbrace$. Time derivatives are defined projections along the fluid four-velocity, and spatial gradients as projections orthogonal to the fluid four-velocity. (Here $\partial_\mu$ refers to a \textit{covariant} derivative.)}
\label{fig:Congruence}
\end{figure*}

In this section, we construct a version of the BGV theorem which does not assume an FRW metric, but relies only on the local properties of the spacetime. Our construction is based on a foliation of the spacetime into spacelike hypersurfaces orthogonal to a congruence of timelike world lines with four-velocity $u^\mu$, as shown in Fig. \ref{fig:Congruence} \cite{Ehlers:1961xww,Ellis:1989jt}. The expansion rate is then defined locally on the orthogonal hypersurfaces, without assumption of a form for the metric, as long as it is consistent with the existence of a timelike congruence $\left\lbrace u^\mu \right\rbrace$. 

Take a timelike geodesic congruence $\left\lbrace u^\mu \right\rbrace$, with metric of signature $(-\ +\ +\ +)$, such that $u^\mu u_\mu = -1$. We can define a coordinate-independent ``time'' derivative
\begin{equation}
    \dot T = \frac{d T}{d s} = u^\mu T_{;\mu},
\end{equation}
where $d s$ is the proper time measured along the world lines of the congruence $\left\lbrace u^\mu \right\rbrace$. The condition for $u^\mu$ to be a geodesic is
\begin{equation}
    \frac{d u^\mu}{d s} = u^\nu u^\mu{}_{;\nu} = 0,
\end{equation}
where a semicolon indicates a covariant derivative. Similarly, we define a gradient operator
\begin{equation}
    \nabla^\mu T = \lambda^{\mu \nu} T_{;\nu},
\end{equation}
where the $\lambda_{\mu \nu}$ is a spatial projection tensor orthogonal to the congruence,
\begin{equation}
    \lambda^{\mu \nu} = g^{\mu \nu} + u^\mu u^\nu,\qquad \lambda^{\mu \nu} u_\nu = 0.
    \label{eq:ProjectionTensor}
\end{equation}
Note especially that $\nabla^\mu$ is a purely spatial gradient operator. 
The expansion rate relative to $\left\lbrace u^\mu \right\rbrace$ is defined by the divergence of the four-velocity
\begin{equation}
    \Theta \equiv u^\mu{}_{;\mu},
\end{equation}
which in a FRW space reduces to the usual Hubble parameter,
\begin{equation}
    \Theta = 3 H. 
\end{equation}
We can write a general covariant derivative of the congruence as
\begin{equation}
    u_{\mu;\nu} = \sigma_{\mu \nu} + \omega_{\mu\nu} + \frac{1}{3} \Theta \lambda_{\mu \nu} - \dot u_\mu u_\nu. 
\end{equation}
For a geodesic congruence, $\dot u_\mu = 0$, with zero shear and vorticity $\sigma = \omega = 0$, we have
\begin{equation}
    u_{\mu;\nu} = \frac{1}{3} \Theta \lambda_{\mu \nu}.
\end{equation}
(We discuss shear in Sec. \ref{sec:Raychaudhuri}.) The spatial gradient of the four-velocity $u$ is then
\begin{equation}
    \nabla^\mu u_\nu = \lambda^{\mu \sigma} u_{\nu;\sigma} = \frac{1}{3} \Theta \lambda^\mu{}_\nu.
\end{equation}
Using $\lambda^{\mu \nu} \lambda_{\mu \nu} = 3$, we have an expression for the expansion rate in terms of the orthogonal gradient of the congruence,
\begin{equation}
    \Theta = \lambda^{\mu \nu} \nabla_\mu u_\nu. 
\end{equation}
We can then define a coordinate- and metric-independent version of $\mathcal{H}_{\mathrm{av}}$ (\ref{eq:Hav}) in terms of the orthogonal gradient as
\begin{equation}
    \Theta_{\mathrm{av}} \equiv \frac{1}{\Delta s} \int{\Theta d s} = \frac{1}{\Delta s} \int{\lambda^{\mu \nu} \nabla_\mu u_\nu \ ds}.
    \label{eq:BasicIntegral}
\end{equation}

We can evaluate the integral (\ref{eq:BasicIntegral}) by noticing that the extrinsic curvature of spatial hypersurfaces orthogonal to the congruence $\left\lbrace u^\mu\right\rbrace$ is
\begin{equation}
    K_{\mu \nu} = \nabla_\mu u_\nu = \frac{1}{2} \mathcal{L}_u \left(\lambda_{\mu \nu}\right),
\end{equation}
where 
\begin{equation}
\mathcal{L}_u \left(\lambda_{\mu \nu}\right) = \lambda_{\mu \sigma} u^{\sigma}{}_{;\nu} + \lambda_{\sigma \nu} u^{\sigma}{}_{;\mu} + \lambda_{\mu \nu; \sigma} u^{\sigma}
\end{equation}
is the Lie derivative along the world line $u^\mu$. The integrand in (\ref{eq:BasicIntegral}) is then 
\begin{equation}
    \lambda^{\mu \nu} \nabla_\mu u_\nu = \frac{1}{2} \lambda^{\mu \nu} \left[\mathcal{L}_u \left(\lambda_{\mu \nu}\right)\right].
\end{equation}
From $\lambda^{\mu \nu} \lambda_{\mu \nu} = 3$, it is tempting to identify $\left(1 / 3\right) \lambda^{\mu \nu}$ as the inverse $\left(\lambda_{\mu \nu}\right)^{-1}$, but the inverse of the projection tensor is undefined, since $\lambda_{\mu \nu}$ is singular, with $\det\left(\lambda\right) = 0$. However, it is straightforward to see that this is an artifact of the fact that $\lambda_{\mu \nu}$ is purely spatial, \textit{i.e.} with one dimension fewer than the full spacetime manifold. In the rest frame of the four-velocity $u^\mu = \left(1, 0, 0, 0\right)$, the definition of the projection tensor (\ref{eq:ProjectionTensor}) tells us that the projection tensor is purely spatial, with the spatial components of the tensor just equal to the corresponding components of the metric:
\begin{equation}
    \lambda_{i j} = g_{i j}.
\end{equation}
In the rest frame of the congruence $\left\lbrace u^\mu\right\rbrace$, the lapse $g_{00} \equiv \mathcal{N} = -1$, and the shift $g_{0 i} = g_{i 0} \equiv \mathcal{N}^i = 0$, by construction. Then $\lambda^{i j} = \left(\lambda_{i j}\right)^{-1}$, and we have the identity
\begin{equation}
    \lambda^{\mu \nu} \left[\mathcal{L}_u \left(\lambda_{\mu \nu}\right)\right] = \lambda^{i j} \left[\frac{d}{d s} \left(\lambda_{i j}\right)\right] = \frac{d}{d s} \mathrm{Tr}\left[\ln{(\lambda_{i j})}\right],
\end{equation}
and we have reduced the integrand in (\ref{eq:BasicIntegral}) to a total derivative. We then have the fully general expression,
\begin{equation}
     \int_{s_i}^{s_f}{\Theta ds} =  \frac{1}{2} \int{\lambda^{\mu \nu} \left[\mathcal{L}_u \left(\lambda_{\mu \nu}\right)\right] ds} = \left.\frac{1}{2} \mathrm{Tr}\left[\ln{(\lambda_{i j})}\right]\right\vert_{s_i}^{s_f}, 
     \label{eq:MainIntegral}
\end{equation}
where $\lambda_{i j}$ are the spatial components of the projection tensor, and the integral is determined by its values on the boundaries at $s_i$ and $s_f$. This is a general version of Eq. (\ref{eq:Hdt}), with the spatial hypersurfaces singular when $\det\left(\lambda_{i j}\right) = 0$.

As an example, we can take a flat FRW metric,
\begin{equation}
    g_{\mu \nu} = \mathrm{diag.}\left[-1, a^2(t), a^2(t), a^2(t)\right].
\end{equation}
Taking $u^\mu$ to be comoving world lines, the projection tensor is then
\begin{equation}
    \lambda_{\mu \nu} = \mathrm{diag.}\left[0, a^2(t), a^2(t), a^2(t)\right],
\end{equation}
and
\begin{equation}
    \frac{1}{2} \lambda^{i j} \left[\mathcal{L}_u \left(\lambda_{i j}\right)\right] =  \frac{3}{2 a^2} \frac{d}{d t} \left(a^2\right) = 3 \frac{\dot a}{a}, 
\end{equation}
which is just the usual Hubble parameter, and we have
\begin{equation}
    \int{\Theta ds} = 3 \int{H dt} = \frac{1}{2} \mathrm{Tr}\left[\ln{(\lambda_{i j})}\right] = 3 \ln{a},
\end{equation}
which is just Eq. (\ref{eq:Hdt}), where the the log of the spatial projection tensor is just log of the scale factor.

We are now in a position to construct a general version of the BGV Theorem. Given a geodesic congruence $\left\lbrace u^\mu \right\rbrace$, choose a second congruence $\left\lbrace v^\mu\right\rbrace \neq \left\lbrace u^\mu\right\rbrace$. We can define
\begin{equation}
    \gamma \equiv - v^\mu u_\mu,
\end{equation}
where $\gamma$ can be identified as the Lorentz boost of the vector $v^\mu$ in the rest frame $u^\mu = (1, 0, 0, 0)$. Since $v^\mu$ is a unit-normalized timelike vector,
\begin{align}
    v^\mu v_\mu &= -1\\
    &= - \gamma^2 + g_{i j} v^i v^j\\
    &= - \gamma^2 + \lambda_{i j} v^i v^j. 
    \label{eq:normalization}
\end{align}
where in the rest frame of $\left\lbrace u^\mu\right\rbrace$, the lapse and shift functions are $\mathcal{N} = -1$ and  $\mathcal{N}^i = 0$, respectively. Normalization of $v^\mu$ results in the relation
\begin{equation}
    \lambda_{i j} v^i v^j = \gamma^2 - 1. 
\end{equation}
We also have the geodesic equation,
\begin{equation}
    \frac{d v^\mu}{d s} = 0,
\end{equation}
where $ds$ is the proper time measured along $v^\mu$. Then
\begin{align}
    \frac{d \gamma}{d s} &= - \frac{d}{d s} \left(v^\mu u_\mu\right)\\
    &= - v^\mu \frac{d u_\mu}{d s} \\
    &= - v^\mu v^\nu u_{\mu;\nu} = - \frac{1}{3} \Theta v^\mu v^\nu \lambda_{\mu \nu} \\
    &= \frac{1}{3} \Theta \left(1 - \gamma^2\right).
\end{align}
where $\left\lbrace u^\mu\right\rbrace$ is shear-free in the rest frame. We then have the relation
\begin{equation}
    \Theta = \frac{3}{1 - \gamma^2} \frac{d \gamma}{d s},
\end{equation}
and
\begin{align}
    \int{\Theta ds} &= \int{\frac{3}{1 - \gamma^2} \frac{d \gamma}{d s} ds} \\
    &= \int{\frac{3 d\gamma}{1 - \gamma^2}} \\
    &= \frac{3}{2} \ln\left(\frac{\gamma + 1}{\gamma - 1}\right).
\end{align}
Therefore, the BGV relation (\ref{eq:BGVRelation}) holds in in any spacetime which admits a local foliation orthogonal to a timelike geodesic congruence, where the expansion is measured by the trace of the log of the spatial projection tensor, $(1/2) \mathrm{Tr}\left[\ln{(\lambda_{i j})}\right]$. As long as the space is locally expanding along a geodesic, the space is past-incomplete. (This proof is equivalent to the original proof in Ref. \cite{Borde:2001nh}, but stated in terms of different variables.) Note that this applies in the inverse as well: if the space is locally \textit{contracting} along a geodesic $\Theta_{\mathrm{av}} < 0$, that geodesic is \textit{future}-incomplete. This will be relevant when we discuss extensions to de Sitter space in Sec. \ref{sec:Raychaudhuri}. 

\section{BGV in Extended de Sitter Spaces}
\label{sec:Raychaudhuri}

\subsection{The Raychaudhuri Equation and coordinates on de Sitter space}

We have seen that the BGV theorem demonstrates geodesic incompleteness of an arbitrary spacetime  by constructing a bounding de Sitter spacetime, such that incompleteness of the bounding spacetime requires incompleteness of the target spacetime, irrespective of the details of its dynamics, as long as the net expansion is positive, with local $\Theta_{\mathrm{av}} > 0$. As such, it is useful to consider the geodesic properties of the bounding de Sitter space. Note that the BGV theorem is a purely \textit{kinematic} statement, relying only on solutions to the geodesic equation, and as such does not rely on any assumption about energy conditions. 

To include the dynamics of the spactime itself, we consider the Raychaudhuri Equation for the local expansion $\Theta$,
\begin{equation}
    \dot\Theta + \frac{1}{3} \Theta^2 + \sigma^2 - \omega^2= -R_{\mu \nu} u^\mu u^\nu,
\end{equation}
where $u^\mu$ is a timelike congruence, $R_{\mu \nu}$ is the Ricci tensor, $\sigma^2$ and $\omega^2$ are the the trace of the shear and vorticity tensors, and the time derivative is defined as a derivative with respect to proper time along the geodesic $u^\mu$. 
\begin{equation}
    \dot\Theta  = \frac{d \Theta}{d s} \equiv u^\mu \Theta_{;\mu}.
\end{equation}
Taking $u^\mu$ to be shear- and vorticity-free, $\sigma^2 = \omega^2 = 0$, de Sitter space is the case of a maximally symmetric spacetime,
\begin{equation}
    \dot\Theta + \frac{1}{3} \Theta^2 = -R_{\mu \nu} u^\mu u^\nu = \Lambda = \mathrm{const.}
\end{equation}
This equation has solutions
\begin{align}
    &\Theta_0 = \pm 3 \sqrt{\frac{\Lambda}{3}},\\
    &\Theta_+ = 3 \sqrt{\frac{\Lambda}{3}} \tanh\left(\sqrt{\frac{\Lambda}{3}} s\right),\\
    &\Theta_- = 3 \sqrt{\frac{\Lambda}{3}} \coth\left(\sqrt{\frac{\Lambda}{3}} s\right).
\end{align}
The solution $\Theta_0$ is easily identified as the standard case of exponential expansion,
\begin{equation}
    H \equiv \frac{\dot a}{a} = \frac{1}{3} \Theta = \sqrt{\frac{\Lambda}{3}} = \mathrm{const.}
\end{equation}
so that
\begin{equation}
    a \propto e^{H t}. 
\end{equation}

The solutions $\Theta_{\pm}$ can be physically interpreted by noting that the Raychaudhuri Equation does not specify the \textit{curvature} of the spacetime. For example, in the case of FRW cosmology, the Raychaudhuri Equation reduces to the usual equation for acceleration, with $\Theta = 3 H$,
\begin{equation}
    \dot\Theta + \frac{1}{3}\Theta^2 = 3 \left(\frac{\ddot a}{a}\right) = \Lambda = \mathrm{const.}
\end{equation}
A complete specification of the spacetime requires the Friedmann Equation,
\begin{equation}
    \left(\frac{\dot a}{a}\right)^2 + \frac{k}{a^2} = \frac{\Lambda}{3} = \mathrm{const.}
\end{equation}
Here $k = 0,\pm 1$ specifies the curvature, and the solution $\Theta_0$ is the solution for a flat ($k = 0$) cosmology,
\begin{equation}
    \left(\frac{\dot a}{a}\right)^2 = \frac{\Lambda}{3} = \mathrm{const.}
\end{equation}
Similarly, the solutions $\Theta_{\pm}$ can be recognized as solutions of the Friedmann Equation for $k = \pm 1$, respectively. In the rest frame of $u^\mu$, $d s = d t$, and taking
\begin{equation}
    a\left(t\right) = \sqrt{\frac{3}{\Lambda}} \cosh\left(\sqrt{\frac{\Lambda}{3}} t\right),\\
\end{equation}
we have
\begin{equation}
    \frac{\dot a}{a} = \sqrt{\frac{\Lambda}{3}} \tanh\left(\sqrt{\frac{\Lambda}{3}} t\right) = \frac{1}{3} \Theta_+,
\end{equation}
and
\begin{align}
    \left(\frac{\dot a}{a}\right)^2 + \frac{1}{a^2} &= \frac{\Lambda}{3} \left[\tanh^2\left(\sqrt{\frac{\Lambda}{3}} t\right) + \frac{1}{\cosh^2}\left(\sqrt{\frac{\Lambda}{3}} t\right)\right] \\
    &= \frac{\Lambda}{3} = \mathrm{const.}
\end{align}
Similarly, for the $\Theta_-$ solution, taking
\begin{equation}
    a\left(t\right) = \sqrt{\frac{3}{\Lambda}} \sinh\left(\sqrt{\frac{\Lambda}{3}} t\right),\\
\end{equation}
we have
\begin{equation}
    \frac{\dot a}{a} = \sqrt{\frac{\Lambda}{3}} \coth\left(\sqrt{\frac{\Lambda}{3}} t\right) = \frac{1}{3} \Theta_-. 
\end{equation}
This then solves the Friedmann Equation with $k = -1$,
\begin{align}
    \left(\frac{\dot a}{a}\right)^2 - \frac{1}{a^2} &= \frac{\Lambda}{3} \left[\coth^2\left(\sqrt{\frac{\Lambda}{3}} t\right) - \frac{1}{\sinh^2}\left(\sqrt{\frac{\Lambda}{3}} t\right)\right] \\
    &= \frac{\Lambda}{3} = \mathrm{const.}
\end{align}

\begin{figure*}
\centerline{\includegraphics[width=6.5in]{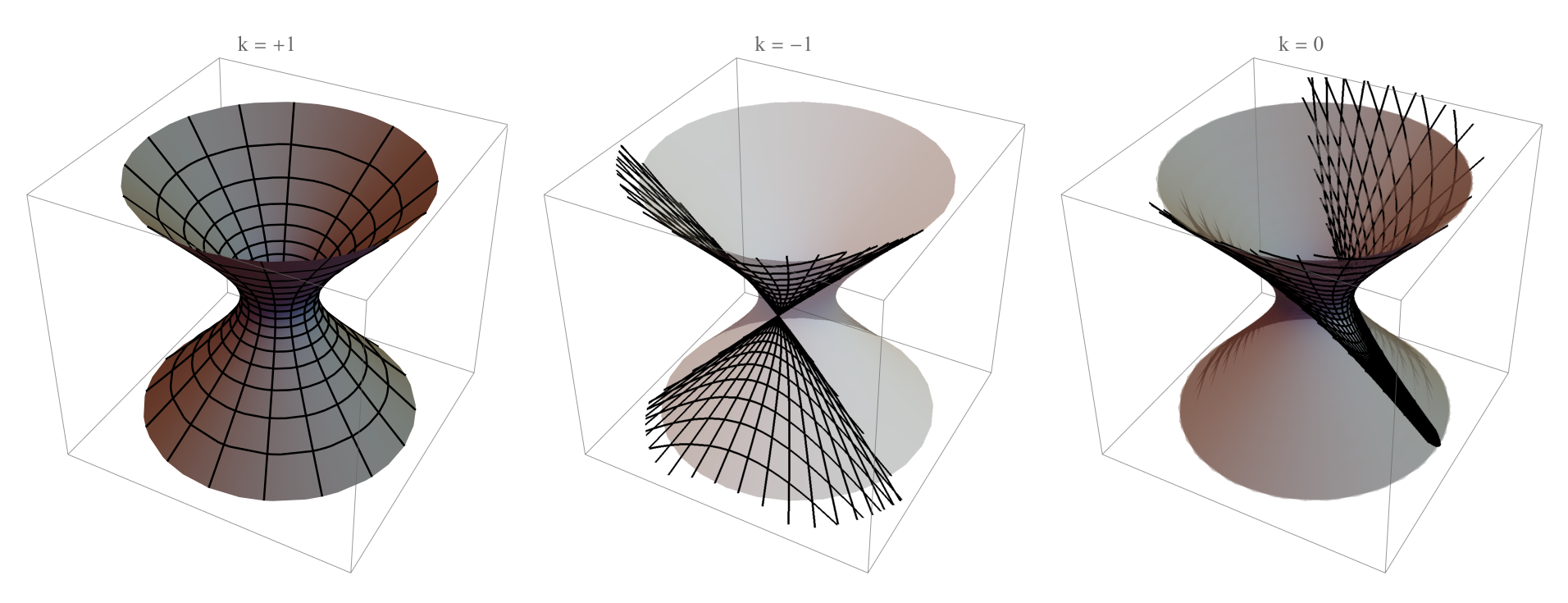}}  
\caption{Closed, flat, and open coordinates (left to right) on the de Sitter invariant hyperboloid.}
\label{fig:dSHyperbola}
\end{figure*}

The solutions $\Theta_{0,\pm}$ can then be seen to correspond to the well-known closed, flat, and open coordinate patches on the complete de Sitter space, defined as the invariant (3+1) hyperboloid in a (4+1) embedding space,
\begin{equation}
    x^2 + y^2 + z^2 + w^2 - v^2 = \alpha^{-2} \equiv \frac{\Lambda}{3} = \mathrm{const.}
\end{equation}
Figure \ref{fig:dSHyperbola} shows the three coordinate patches on the invariant hyperbola. The $k = +1$ coordinates cover the entire hyperbola, corresponding to a closed universe with metric
\begin{equation}
    ds^2 = -dt^2 + \alpha^2 \cosh^2\left(\frac{t}{\alpha}\right) \left[d\chi^2 + \sin^2\left(\chi\right)\left(d\theta^2 + \sin^2{\theta} d\phi^2\right)\right].
    \label{eq:closedmetric}
\end{equation}
In this patch, the curvature radius $a$ which contracts from $t \rightarrow -\infty$ to finite radius at $a = \alpha^{2} = 3 / \Lambda$ at $t = 0$, then expands outward again as $t \rightarrow \infty$.

The $k = 0, -1$ coordinates, by contrast, cover only parts of the hyperbola. The $k = -1$ coordinates correspond to an open universe which expands out of an initial singularity at $t = 0$, which appears on the full manifold as a bubble nucleation. A dual coordinate patch is an ``antibubble'', collapsing from $t \rightarrow - \infty$ to a singularity at $t = 0$. The metric in the bubble/antibubble regions is
\begin{equation}
    ds^2 = -dt^2 + \alpha^2 \sinh^2\left(\frac{t}{\alpha}\right) \left[d\chi^2 + \sinh^2\left(\chi\right)\left(d\theta^2 + \sin^2{\theta} d\phi^2\right)\right].
    \label{eq:openmetric}
\end{equation}
The bubble is geodesically past-incomplete, $s = [0,\infty]$, and the antibubble is future-incomplete, $s = [-\infty, 0]$. 

\begin{figure*}
\centerline{\includegraphics[width=6.5in]{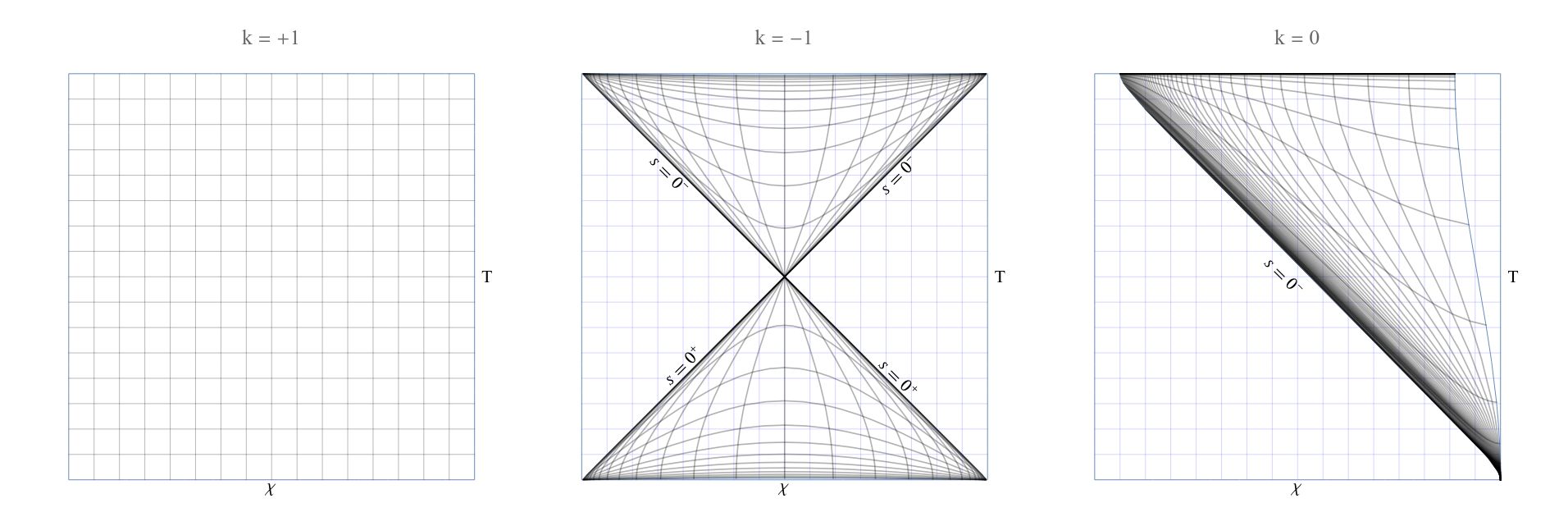}}  
\caption{Penrose diagrams for closed, flat, and open coordinates (left to right) on de Sitter space. The $k = +1$ coordinates cover the entire space, and are nonsingular. The $k = -1$ coordinates consist of bubble/antibubble solutions, connected via a singularity at $t = 0$. The $k = 0$ coordinates cover the upper half-diagonal, evolving out of singularity at $t \rightarrow -\infty$. Singularities are labeled by $s = 0^{\pm}$. }
\label{fig:dSPenrose}
\end{figure*}

The $k = 0$ patch is the flat universe, and covers the upper half of the hyperbola, expanding exponentially from a singularity at $t \rightarrow -\infty$, extending forward in time to $t \rightarrow +\infty$, with metric
\begin{equation}
    ds^2 = -dt^2 + \alpha^2 e^{2 t / \alpha} \left[d\chi^2 + \chi^2 \left( d\theta^2 + \sin^2{\theta} d\phi^2\right)\right].
    \label{eq:flatmetric}
\end{equation}
Penrose diagrams for the three coordinate patches are shown in Fig. \ref{fig:dSPenrose}. (The Appendix details the specific coordinate transformations used to construct the plots.) While the $k = +1$ coordinates are nonsingular, the $k = -1$ and $k = 0$ patches contain singularities in the spatial projection tensor, $\det\left(\lambda_{i j}\right) \rightarrow 0$, corresponding to $a \rightarrow 0$ in FRW coordinates, where the intrinsic curvature of the spatial hypersurfaces diverges. The four-dimensional Ricci scalar $\mathcal{R}$, however, is everywhere \textit{finite}, with
\begin{equation}
\mathcal{R} = 4 \Lambda. 
\end{equation}
In this sense, the singularities in the flat and open coordinate systems on the de Sitter manifold are coordinate, rather than physical singularities, as long as the de Sitter symmetry is exact.

\subsection{Coordinates on de Sitter space and the BGV Theorem}

To make contact with the BGV Theorem, take $u^\mu$ to be the comoving four-velocity in the flat patch $\Theta_0$, and consider non-comoving timelike geodesics $v^\mu$ satisfying Eq. (\ref{eq:dtdstimelike}), 
\begin{equation}
    v^\mu = \left(\gamma, \frac{1}{a} \sqrt{\gamma^2 -1} \vec{n}\right) = \left(\sqrt{1 + \frac{v_0^2}{a^2}}, \frac{v_0}{a^2} \vec{n}\right),
    \label{eq:vcongruence}
\end{equation}
where $v_0 = \mathrm{const.}$ and $\vec{n} \cdot \vec{n} = 1$. Then
\begin{equation}
    g_{\mu \nu} v^\mu v^\nu = -\gamma^2 + a^2 \left(\frac{\gamma^2 - 1}{a^2}\right) = -1.  
\end{equation}
It is straightforward to identify the geodesics $v^\mu$ as rest-frame geodesics in the open patch $\Theta_-$ by taking $v_0 = \alpha = \sqrt{3 / \Lambda}$, and
\begin{equation}
    a\left(s\right) = \alpha \sinh\left(\alpha^{-1} s\right),
\end{equation}
where $s$ is the proper time along the geodesic $v^\mu$. Then
\begin{equation}
    \sqrt{1 + \frac{v_0^2}{a^2}} = \frac{\cosh\left(\alpha^{-1} s\right)}{\sinh\left(\alpha^{-1} s\right)},
\end{equation}
and
\begin{equation}
    \frac{v_0}{a^2} = \frac{\alpha^{-1}}{\sinh^2\left(\alpha^{-1} s\right)},
\end{equation}
so that
\begin{equation}
    g_{\mu \nu} v^\mu v^\nu = - \frac{\cosh^2\left(\alpha^{-1} s\right)}{\sinh^2\left(\alpha^{-1}s\right)} + \alpha^{2} \sinh^2\left(\alpha^{-1} s \right) \left[\frac{\alpha^{-1}}{\sinh^2\left(\alpha^{-1} s\right)}\right]^2 \left(\vec{n} \cdot \vec{n}\right) = -1.
\end{equation}
This makes clear the physical interpretation of the geodesics in the bounding de Sitter space used to prove the BGV theorem. While comoving observers $\left\lbrace u^\mu\right\rbrace$ see themselves in a universe with a flat FRW metric (\ref{eq:flatmetric}), observers on non-comoving timelike geodesics $\left\lbrace v^\mu\right\rbrace$ see, in their rest frame, the metric for an \textit{open} universe (\ref{eq:openmetric}), which is trivially past-incomplete, having a singularity at $s = 0$. This can also be seen by noting that in the rest frame of the non-comoving observer, the congruence $\left\lbrace v^\mu\right\rbrace$ is shear-free, while in the comoving rest-frame, the congruence $\left\lbrace v^\mu\right\rbrace$ (\ref{eq:vcongruence}) has shear 
\begin{eqnarray}
    \sigma^2 &&= \sigma^{\mu \nu} \sigma_{\mu \nu} = \frac{2}{3} \left(\frac{a'}{a}\right)^2 \frac{v_0^4}{a^4 \left(1 + v_0^2 / a^2\right)} = \left(\frac{2}{3 \alpha^2}\right) \frac{\left(\gamma^2 - 1\right)^2}{\gamma^2},
\end{eqnarray}
so that curvature in the rest frame of $\left\lbrace v^\mu\right\rbrace$ manifests as shear in the rest frame of $\left\lbrace u^\mu\right\rbrace$. (Direct coordinate transformations between the $k = 0,\pm 1$ coordinate patches on the de Sitter manifold are discussed in the Appendix.)

We can similarly see that geodesics $\left\lbrace v^\mu\right\rbrace$ in the rest frame of the \textit{closed} patch, with metric (\ref{eq:closedmetric}), are \textit{spacelike} in the comoving rest frame $\left\lbrace u^\mu\right\rbrace$ defined on the flat patch,
\begin{equation}
    v^\mu v_\mu = -\left(\frac{dt}{ds}\right)^2 + a^2 \left\vert \frac{d \mathbf{x}}{d t}\right\vert^2 = +1,
\end{equation}
with geodesic equation
\begin{equation}
    a^2 \left\vert  \frac{d \mathbf{x}}{d t}\right\vert = v_0 = \mathrm{const.}
\end{equation}
Then
\begin{equation}
    \gamma = \frac{dt}{ds} = \sqrt{\frac{v_0^2}{a^2} - 1},
\end{equation}
and 
\begin{equation}
    v^\mu = \left(\gamma, \frac{1}{a} \sqrt{\gamma^2 -1} \vec{n}\right) = \left(\sqrt{\frac{v_0^2}{a^2} - 1}, \frac{v_0}{a^2} \vec{n}\right).
\end{equation}
If we identify the metric $a\left(s\right)$ with the metric for the closed patch (\ref{eq:closedmetric}),
\begin{equation}
    a\left(s\right) = \alpha \cosh\left(\alpha^{-1} s\right),
\end{equation}
then
\begin{equation}
    \sqrt{\frac{v_0^2}{a^2} - 1} = \sqrt{- \tanh^2\left(\alpha^{-1} s\right)},
\end{equation}
and
\begin{equation}
    g_{\mu \nu} v^\mu v^\nu = \tanh^2\left(\alpha^{-1} s\right) + \frac{1}{\cosh^2\left(\alpha^{-1} s\right)} \left(\vec{n} \cdot \vec{n}\right) = +1.
\end{equation}
This case is of less physical interest, since the closed-patch geodesics are tachyonic in the flat patch, and we will not consider it further. 

\subsection{Application to extended de Sitter spaces}

By considering solutions to the Raychaudhuri Equation and the corresponding coordinate patches on a full de Sitter manifold, we have arrived at a straightforward geometric picture of the construction used in the BGV theorem, which states that spacetimes with net positive expansion $\Theta_{\mathrm{av}} > 0$ are geodesically past-incomplete, and spacetimes with net negative expansion $\Theta_{\mathrm{av}} < 0$ are geodesically future-incomplete. This is shown by constructing a bounding de Sitter space with the same average expansion rate $\Theta = \Theta_{\mathrm{av}} = \mathrm{const.}$ as the target space. In the bounding space, a comoving observer in the (expanding) flat coordinate patch sees a singularity $\det\left(\lambda_{i j}\right) \rightarrow 0$ only at $t \rightarrow -\infty$, where $\lambda_{i j}$ is the spatial projection tensor orthogonal to $u^\mu$. However, there also exists a second set of well-defined timelike geodesics $v^\mu$, such that the rest frame of $v^\mu$ is the open patch on the de Sitter manifold. In this patch, the singularity $\det\left(\lambda_{i j}\right) \rightarrow 0$ occurs at finite past time, $s = 0$, and the space is therefore geodesically past-incomplete. 

With this geometric picture of the BGV construction in hand, we now turn to the case of extended manifolds such as the Conformal Cyclic Cosmology (CCC), proposed by Penrose \cite{Penrose:2010}. In Penrose's construction, the asymptotic $t \rightarrow \infty$ boundary of one expanding universe (dubbed an \textit{aeon}), is matched to the asymptotic $t \rightarrow - \infty$ boundary of a subsequent aeon, with the the metric conformally rescaled at the matching
\begin{equation}
    \hat{g}_{\mu \nu} = \hat{\Omega}^2 g_{\mu \nu} = \check{\Omega}^2 g_{\mu \nu} = \check{g}_{\mu \nu},
    \label{eq:ConformalMapping}
\end{equation}
where $\hat{g}_{\mu \nu}$ is the metric in the future aeon, and $\check{g}_{\mu \nu}$ is the metric in the past aeon, connected via a bridging metric $g_{\mu \nu}$ and conformal scaling factors $\Omega$. A number of realizations of this (somewhat vague) prescription have been proposed. In particular, the exact nature of the conformal rescaling at the boundary is unspecified, and different conventions are possible. Ref. \cite{Stevens:2022hae} provides a useful review of various proposals. 

\begin{figure*}
\centerline{\includegraphics[width=3.0in]{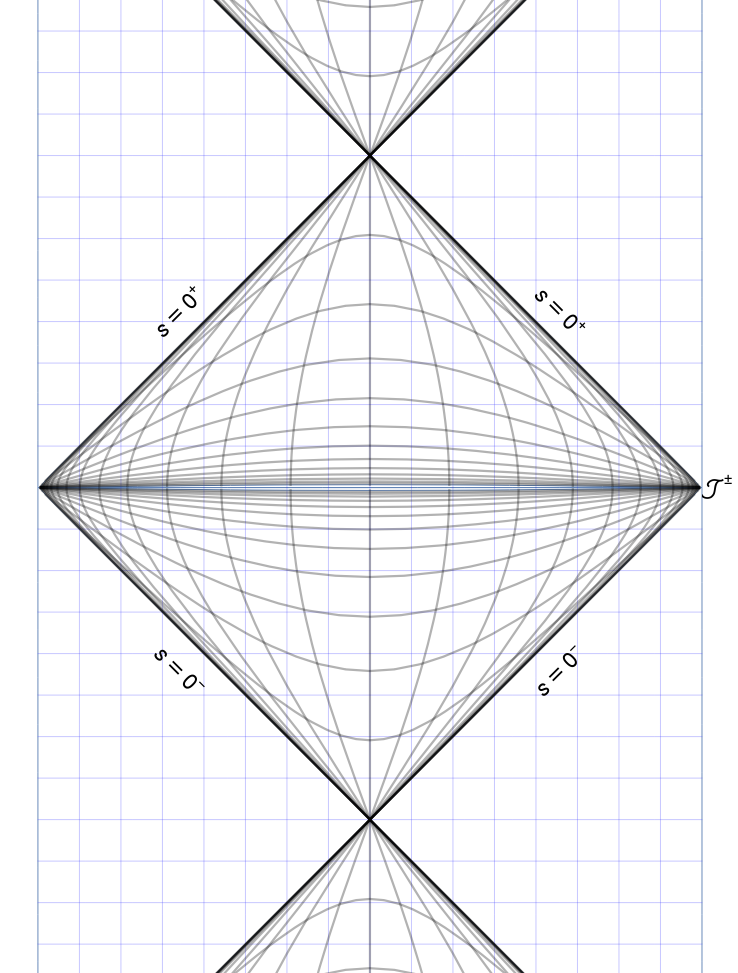}}  
\caption{Matching of asymptotically de Sitter spaces at $\mathcal{J}^\pm$. The $\mathcal{J}^+$ asymptote of the geodesically past-incomplete bubble coordinates map to the $\mathcal{J}^-$ coordinates of the geodesically future-incomplete antibubble coordinates. Past singularities are at $s = 0^-$, and future singularities are at $s = 0^+$.}
\label{fig:CCC}
\end{figure*}

Here we consider Penrose's matching construction on manifolds relevant to realistic cosmological models, dominated at late time by a cosmological constant. In this case, the asymptotic behavior of the spacetime is de Sitter, and we can construct a bounding de Sitter space on each aeon exactly as in the usual BGV theorem in Sec. \ref{sec:Raychaudhuri}. In this case, matching between aeons will be at the null boundaries $\mathcal{J}^\pm$ of successive aeons. We focus on the $k = -1$ bubble/antibubble coordinates, relevant for the BGV construction. The matching at $\mathcal{J}^\pm$ maps the (expanding) bubble coordinates in the past aeon onto the (contracting) antibubble coordinates in the future aeon, with the geodesics spanning the aeons continuous across the boundary, as shown in Fig. \ref{fig:CCC}. By symmetry, the net expansion from the past bubble boundary $s = 0^-$ to the future antibubble boundary $s = 0^+$, vanishes,
\begin{equation}
    \int_{s = 0^-}^{s = 0^+}{\Theta ds} = 0,
\end{equation}
and the geodesics extending between the past and future singularities are of infinite proper length. However, any physical observer in the expanding bubble coordinates of the past aeon will see themselves on a geodesic which is \textit{finite} to the past, and any physical observer in the contracting antibubble coordinates of the future aeon will see themselves on a geodesic which is finite to the future, with both ending in singularities at proper time coordinate $s = 0^\pm$. The only observers who will see themselves on geodesics which are both past- and future-complete are the observers at the asymptotic boundary $\mathcal{J}^\pm$. In this sense, the cyclic extension of a de Sitter manifold remains geodesically incomplete. Note especially that this conclusion depends \textit{only} on the geodesic structure of the spacetime, and is independent of the details of how the metric is conformally rescaled at the boundary, Eq. (\ref{eq:ConformalMapping}). 

Finally, we comment on the nature of the singularities at $s = 0^{\pm}$. In the exact de Sitter space, these are coordinate singularities $\det\left(\lambda_{i j}\right) \rightarrow 0$, not physical singularities, since the Ricci curvature $\mathcal{R}$ remains finite at the singularity, and the space can be spanned by the (non-singular) closed coordinates, as pointed out by Aguirre and Gratton \cite{Aguirre:2001ks}, who argue that this means that the geodesically incomplete open patch can be consistently extended without the presence of a physical singularity. However, this is true only of the \textit{bounding} space, which is protected by the presence of an exact de Sitter symmetry. The real universe -- the target space -- contains matter and radiation, which break the de Sitter symmetry that protects the Ricci curvature at $s = 0^\pm$, so that even inflationary spacetimes are at best asymptotically de Sitter on the past boundary. The extendibility of such quasi-de Sitter spacetimes onto geodesically complete manifolds is discussed in detail by Geshnizjani, \textit{et al.} in Ref. \cite{Geshnizjani:2023edw}, who argue that physical singularities are generic in the absence of very special boundary conditions.

\section{Conclusions}
\label{sec:Conclusions}

In this paper, we have derived a general version of the geodesic incompleteness theorem of Borde, Guth, and Vilenkin \cite{Borde:2001nh}, which states that any spacetime with local net positive expansion on an interval will be geodesically incomplete on that interval. Our derivation is in many ways similar to that of Kothwala in Ref. \cite{Kothawala:2018ghr}, except instead of defining the expansion in terms of directly constructed orthogonal deviation vectors between geodesics, we instead quantify it using the spatial projection tensor $\lambda_{\mu \nu}$ orthogonal to geodesic flow lines, showing that the expansion rate $\Theta$ integrated over proper time $ds$ along a geodesic is determined by the trace of the log of the spatial projection tensor on the boundary,
\begin{equation}
     \int_{s_i}^{s_f}{\Theta ds}  = \left.\frac{1}{2} \mathrm{Tr}\left[\ln{(\lambda_{i j})}\right]\right\vert_{s_i}^{s_f}. 
\end{equation}

We reproduce the general proof of the BGV theorem using this formalism, which is purely a kinematic argument, and does not rely on assumptions about energy conditions, only that the average expansion rate $\Theta_{\mathrm{av}}$ over an interval be positive. The BGV theorem relies on constructing a bounding de Sitter space with expansion rate $\Theta_{\mathrm{av}}$, such that geodesic incompleteness of the bounding space implies geodesic incompleteness of the target space. We apply the Raychaudhuri equation to study the geodesic structure of the bounding de Sitter space, identifying incomplete geodesics with rest-frame observers in the open patch on the invariant hyperboloid. Geodesics which are non-comoving in the rest frame of a flat ($k = 0$) exponentially expanding FRW coordinate system define a foliation that is characterized by a negative curvature ($k = -1$) FRW metric, such that the interval $t = [-\infty,\infty]$ on the flat coordinates maps to $s = [0, \infty]$ on the open coordinates, with a singularity at finite past proper time. In this sense, the BGV construction is a mapping between flat and open coordinates on the full de Sitter hyperboloid, and we demonstrate this correspondence by construction. 

Finally, we apply the result to the extension suggested by Penrose's Conformal Cyclic Cosmology, which proposes constructing cyclic cosmology by matching ``aeons'' at surfaces at future infinity $\mathcal{J}^{\pm}$, with a conformal rescaling at each matching. We apply the matching prescription to asymptotically de Sitter cosmologies, and consider the geodesic structure of the extended space, which is independent of the details of the conformal rescaling. We show that the geodesically past-incomplete open patch on de Sitter extends to a sequence of bubble/antibubble regions, connected by singularities, where the (expanding) bubble regions are geodesically past-incomplete and the (contracting) antibubble regions are geodesically future-incomplete. We discuss the application to physically realistic cosmologies, and argue that these boundaries represent physical singularities in the absence of an exact de Sitter symmetry \cite{Geshnizjani:2023edw}.

\section*{Acknowledgements}

This work is supported by the National Science Foundation under grant NSF-PHY-2014021. WHK thanks the Indian Insitute of Technology, Madras for hospitality while a portion of this work was being completed. 
SM would like to thank the Indian Institute of Technology Madras, Chennai, 
India, for support through the Half-Time Research Assistantship. 
We thank Dawood Kothawala for helpful conversations. 

\vfill
\section*{Appendix: Coordinate systems in de Sitter Space}

Here, we detail the coordinate transformations used to construct the $k = 0, \pm 1$ patches on de Sitter space shown in Fig. \ref{fig:dSHyperbola}, and to construct the Penrose diagrams in Fig. \ref{fig:dSPenrose} \cite{Hawking:1973uf}. We begin by constructing a static hyperbola in a $4+1$-dimensional space, with coordinates $(x, y, z, w, v)$, with metric
\begin{equation}
    ds_5^2 = dx^2 + dy^2 + dz^2 + dw^2 - dv^2. 
\end{equation}
We construct de Sitter space as an invariant hyperbololoid, defined as the surface
\begin{equation}
    x^2 + y^2 + z^2 + w^2 - v^2 = \frac{\Lambda}{3}. 
\end{equation}
This is a $(3+1)$-dimensional surface in the $(4+1)$-dimensional embedding manifold. 

The closed coordinates on the hyperbola corresponding to the $\Theta_+$ solution can be constructed by defining a coordinate system $(t, \chi, \theta, \phi)$ as follows:
\begin{align}
    v &= \alpha \sinh\left(t / \alpha\right),\\
    w &= \alpha \cosh\left(t / \alpha\right) \cos\left(\chi\right),\\
    x &= \alpha \cosh\left(t / \alpha\right) \sin\left(\chi\right) \cos\left(\theta\right),\\
    y &= \alpha \cosh\left(t / \alpha\right) \sin\left(\chi\right) \sin\left(\theta\right) \cos\left(\phi\right),\\
    z &= \alpha \cosh\left(t / \alpha\right) \sin\left(\chi\right) \sin\left(\theta\right) \sin\left(\phi\right).
    \label{eq:closed}
\end{align}
Here
\begin{equation}
    \frac{\Lambda}{3} \equiv \alpha^{-1}. 
\end{equation}
The metric on the coordinates $(t, \chi, \theta, \phi)$ is of the $k = +1$ FRW form,
\begin{equation}
    ds^2 = -dt^2 + \alpha^2 \cosh^2\left(\frac{t}{\alpha}\right) \left[d\chi^2 + \sin^2\left(\chi\right)\left(d\theta^2 + \sin^2{\theta} d\phi^2\right)\right].
\end{equation}

The open coordinates on the hyperbola corresponding to the $\Theta_-$ solution can be constructed by defining a coordinate system $(t, \chi, \theta, \phi)$ as follows:
\begin{align}
    v &= \alpha \sinh\left(t / \alpha\right) \cosh\left(\chi\right),\\
    w &= \alpha \cosh\left(t / \alpha\right),\\
    x &= \alpha \sinh\left(t / \alpha\right) \sinh\left(\chi\right) \cos\left(\theta\right),\\
    y &= \alpha \sinh\left(t / \alpha\right) \sinh\left(\chi\right) \sin\left(\theta\right) \cos\left(\phi\right),\\
    z &= \alpha \sinh\left(t / \alpha\right) \sinh\left(\chi\right) \sin\left(\theta\right) \sin\left(\phi\right).
    \label{eq:open}
\end{align}
The metric on the coordinates $(t, \chi, \theta, \phi)$ is of the $k = -1$ FRW form,
\begin{equation}
    ds^2 = -dt^2 + \alpha^2 \sinh^2\left(\frac{t}{\alpha}\right) \left[d\chi^2 + \sinh^2\left(\chi\right)\left(d\theta^2 + \sin^2{\theta} d\phi^2\right)\right].
\end{equation}

The flat coordinates on the hyperbola corresponding to the $\Theta_0$ solution can be constructed by defining a coordinate system $(t, \chi, \theta, \phi)$ as follows:
\begin{align}
    v &= \alpha \sinh\left(t / \alpha\right) + \frac{\alpha}{2} \chi^2 e^{t / \alpha},\\
    w &= \alpha \cosh\left(t / \alpha\right) - \frac{\alpha}{2} \chi^2 e^{t / \alpha},\\
    x &= \alpha \chi e^{t / \alpha} \cos\left(\theta\right),\\
    y &= \alpha \chi e^{t / \alpha} \sin\left(\theta\right) \cos\left(\phi\right),\\
    z &= \alpha \chi e^{t / \alpha} \sin\left(\theta\right) \sin\left(\phi\right).
    \label{eq:flat}
\end{align}
The metric on the coordinates $(t, \chi, \theta, \phi)$ is of the $k = 0$ FRW form,
\begin{equation}
    ds^2 = -dt^2 + \alpha^2 e^{2 t / \alpha} \left[d\chi^2 + \chi^2 \left( d\theta^2 + \sin^2{\theta} d\phi^2\right)\right].
\end{equation}

To construct the Penrose diagrams in Fig. \ref{fig:dSPenrose}, we begin with the closed patch, which is especially simple. We compactify the time dimension from $t = \left[-\infty,\infty\right] \rightarrow T = \left[-\pi/2,\pi/2\right]$ by the transformation
\begin{equation}
    T \equiv 2 \arctan\left(e^{t/\alpha}\right) - \frac{\pi}{2}.
    \label{eq:Tcompactification}
\end{equation}
The metric is then conformally equivalent to spherical coordinates in Minkowski space, 
\begin{equation}
    d S^2 =  \alpha^2 \sec^2\left({T}{}\right) \left[-d T^2 + d\chi^2 + \sin^2\left(\chi\right)\left(d\theta^2 + \sin^2{\theta} d\phi^2\right)\right],
\end{equation}
with the coordinates $T$ and $\chi$ spanning
\begin{align}
    T &= \left[-\frac{\pi}{2},\frac{\pi}{2}\right],\\
    \chi &= \left[-\frac{\pi}{2},\frac{\pi}{2}\right].
\end{align}

To construct the Penrose diagram for the $k = -1$ case, we first map the coordinates $\bar{t}, \bar{\chi}$ on the open patch Eq. (\ref{eq:open}) to the coordinates $t, \chi$ on the closed patch, Eq. (\ref{eq:closed}), 
\begin{align}
    t &= \alpha\, \mathrm{arcsinh}\left[\sinh\left(\frac{\bar{t}}{\alpha}\right) \cosh\left(\bar{\chi}\right)\right],\\
    \chi &= \arctan\left[\tanh\left(\frac{\bar{t}}{\alpha}\right) \sinh\left(\bar{\chi}\right)\right],
\end{align}
and then compactify the time dimension using Eq.~(\ref{eq:Tcompactification}),
\begin{equation}
    T \equiv 2 \arctan\left(e^{t/\alpha}\right) - \frac{\pi}{2}
\end{equation}

To construct the Penrose diagram for the $k = 0$ case, we first map the coordinates $\bar{t}, \bar{\chi}$ on the open patch Eq. (\ref{eq:open}) to coordinates on the invariant hyperbola
\begin{align}
    v &= \alpha \sinh\left(\bar{t} / \alpha\right) + \frac{\alpha}{2} \bar{\chi}^2 e^{\bar{t} / \alpha},\\
    w &= \alpha \cosh\left(\bar{t} / \alpha\right) - \frac{\alpha}{2} \bar{\chi}^2 e^{\bar{t} / \alpha},\\
    r &= \alpha \chi e^{\bar{t} / \alpha},
\end{align}
where we have suppressed the angular coordinates by defining $r^2 = x^2 + y^2 + z^2$. We then map the coordinates $v, w, r$ onto the closed patch by the transformation
\begin{align}
    t &= \alpha \,\mathrm{arcsinh}\left(\frac{v}{\alpha}\right),\\
    \chi &= \arctan\left(\frac{w}{r}\right).
\end{align}
and then compactify the time dimension $t$ using Eq. (\ref{eq:Tcompactification}).

\bibliography{Paper.bib}
\end{document}